\documentclass{article}%
\usepackage{amsfonts}
\usepackage{amsmath}
\usepackage{amssymb}
\usepackage[brazilian]{babel}
\usepackage{graphicx}
\usepackage{geometry}
\usepackage[justification=raggedright]{caption}%
\providecommand{\U}[1]{\protect\rule{.1in}{.1in}}
\newtheorem{theorem}{Theorem}

\newtheorem{definition}[theorem]{Definition}
\newtheorem{example}[theorem]{Example}

\begin{document}
\pagestyle{myheadings} \markboth{\footnotesize Rodrigues}{\footnotesize
Introdu\c{c}\~{a}o as t\'{e}cnicas do c\'{a}lculo fracion\'{a}%
rio para estudar modelos da
f\'{i}sica matem\'{a}tica}
\title{Introdu\c{c}\~{a}o \`{a}s t\'{e}cnicas do c\'{a}lculo fracion\'{a}%
rio para estudar modelos da
f\'{i}sica matem\'{a}tica\\\small{\textit
{(Introduction to the techniques of the fractional calculus to investigate
some models of the mathematical physics)}}}
\date{}
\maketitle{}
\begin{center}
\large{Fabio G. Rodrigues\footnote{E-mail: fabior@mpcnet.com.br\\
}}, {E. C. de Oliveria\footnote{E-mail:capelas@ime.unicamp.br\\
}}
\end{center}
\begin{center}
\small{\textit{${}^{1,2}$Instituto de Matem\'{a}tica, Estat\'{i}%
stica e Computa\c{c}\~{a}o Cient\'{i}%
fica, Universidade Estadual de Campinas, Campinas, SP, Brasil}}\\
\end{center}
\thispagestyle{empty}
	
\begin{abstract}
{\small
Neste trabalho, recorremos \`{a} metodologia da transformada de Laplace a fim de mostrar a sua import\^{a}ncia na abordagem de uma classe de equa\c{c}\~{o}es diferenciais fracion\'{a}rias. Em particular, apresentamos aplica\c{c}\~{o}es desta metodologia ao discutirmos poss\'{i}veis generaliza\c{c}\~{o}es de certos problemas f\'{i}sicos no campo da viscoelasticidade linear e osciladores harm\^{o}nicos, comprovando que o uso do c\'{a}lculo fracion\'{a}rio em modelagem e resolu\c{c}\~{o}o de problemas usualmente abordados pelo c\'{a}lculo de ordem inteira, oferece vantagens promissoras para nos fornecer formula\c{c}\~{o}es mais consistentes com os dados experimentais.\\
\textbf{Palavras-chave:}
C\'{a}lculo fracion\'{a}rio, Equa\c{c}\~{o}es diferenciais fracion\'{a}rias; Transformada de Laplace.\\
	
In this paper, we resort to the Laplace transform method in order to show its efficiency when approaching some types of fractional differential equations. In particular, we present some applications of such methods when applied to possible generalizations of certain physical problems in linear viscoelasticity and harmonic oscillators, proving that fractional calculus is well suited for the modelling and solving of problems usually treated by ordinary integer calculus, with the promissing advantages of being able to provide more accurate theoretical predictions to fit with experimental data.
}
\end{abstract}
\normalsize\baselineskip=12pt%

\section{Introdu\c{c}\~{a}o}

O c\'{a}lculo de ordem n\~{a}o inteira, popularmente conhecido como
c\'{a}lculo fracional ou C\'{a}lculo Fracion\'{a}rio (CF), de uma maneira
simples, tem a inten\c{c}\~{a}o de generalizar o c\'{a}lculo integral e
diferencial, conforme proposto, independentemente, por Newton e Leibniz. Aqui,
evitamos, sempre que poss\'{\i}vel, o aparato matem\'{a}tico envolvendo
explicitamente as f\'{o}rmulas advindas, por exemplo, das diversas maneiras de
calcular uma derivada. Ainda mais, utiliza-se a nomenclatura CF por entender
que o nome est\'{a} totalmente consolidado e, em l\'{\i}ngua portuguesa, \'{e}
uma tradu\c{c}\~{a}o livre de \emph{fractional calculus}. Um estudo versando
sobre a linha do tempo envolvendo o CF pode ser encontrada em \cite{Tenreiro2,
Tenreiro3, Tenreiro4} enquanto um cap\'{\i}tulo sobre a hist\'{o}ria do
c\'{a}lculo fracion\'{a}rio pode ser encontrada em \cite{TeseHeron, RubensECO}.

Como j\'{a} mencionado, existe mais de uma maneira de calcular a derivada e,
portanto, parece eminente quest\~{o}es do tipo: Para que serve o CF? Onde
utiliz\'{a}-lo? Existe uma interpreta\c{c}\~{a}o geom\'{e}trica e/ou
f\'{\i}sica? Qual a rela\c{c}\~{a}o, se \'{e} que existe, com o c\'{a}lculo de
ordem inteira? Muitas outras quest\~{o}es de ordem matem\'{a}tica ou de ordem
f\'{\i}sica podem ser colocadas. Por exemplo, apenas para mencionar duas
dessas quest\~{o}es: Ser\'{a} que existe uma regra da cadeia associada \`{a}
derivada de ordem fracion\'{a}ria? E o teorema fundamental do c\'{a}lculo,
teorema que coroa os c\'{a}lculos diferencial e integral, tem um an\'{a}logo
fracion\'{a}rio? Ainda mais, no c\'{a}lculo de ordem inteira emerge uma classe
de fun\c{c}\~{o}es, as chamadas fun\c{c}\~{o}es especiais, solu\c{c}\~{o}es
das equa\c{c}\~{o}es diferenciais ordin\'{a}rias e/ou parciais que descrevem
um particular sistema, enquanto no CF emerge tamb\'{e}m uma classe de
fun\c{c}\~{o}es a ele associado, solu\c{c}\~{a}o de um particular problema
cuja derivada \'{e} de ordem n\~{a}o inteira, as fun\c{c}\~{o}es especiais do
CF \cite{Herrmann, Hilfer, Kilbas, Mainardi, Miller, Hammond, Oldham,
Tenreiro1, Podlubny, Podlubny1, Podlubny2, Gorenflo, Samko, Eliana, TeseFabio}.

O objetivo principal desse trabalho \'{e} introduzir as ferramentas
b\'{a}sicas a fim de discutir um particular problema, composto por
equa\c{c}\~{a}o diferencial fracion\'{a}ria e condi\c{c}\~{a}o inicial. De
modo a atingir este objetivo, isto \'{e}, discuss\~{a}o e resolu\c{c}\~{a}o de
uma equa\c{c}\~{a}o diferencial fracion\'{a}ria, aborda-se o tema derivada
fracion\'{a}ria que, por sua vez, requer o conceito de integral
fracion\'{a}ria. De uma maneira bastante simplificada, introduzimos o conceito
de integral fracion\'{a}ria para depois, utilizando tal conceito, introduzir a
derivada fracion\'{a}ria que, nesse trabalho discute apenas as
formula\c{c}\~{o}es conforme propostas por Riemann-Liouville e por Caputo
\cite{Herrmann, Kilbas, Mainardi, Miller, Oldham, Podlubny, Samko, Capelas,
Kai}. Em particular, apesar de mais restritiva, a derivada de Caputo admite a
mesma interpreta\c{c}\~{a}o para as condi\c{c}\~{o}es iniciais que a
formula\c{c}\~{a}o cl\'{a}ssica de ordem inteira \cite{Hammond, Podlubny,
Podlubny1}. Mostrou-se tamb\'{e}m que essas duas formula\c{c}\~{o}es podem ser
recuperadas a partir da defini\c{c}\~{a}o de Gr\"{u}nwald-Letnikov
\cite{TeseFabio}.

De uma maneira simples e objetiva, pode-se pensar nos operadores de ordem
fracion\'{a}ria como os operadores que representam fun\c{c}\~{o}es da
mem\'{o}ria sobre a hist\'{o}ria de alguns sinais de sistemas f\'{\i}sicos.
Por exemplo, uma integral de primeira ordem, de uma vari\'{a}vel que
representa o estado de um sistema, pode ser pensada como uma soma em que se
atribuem pesos a cada ponto, com todos os pontos ponderados com o mesmo peso,
independentemente de qu\~{a}o longe est\~{a}o no passado. Isso sobre toda a
hist\'{o}ria do sistema. Uma integral de ordem fracion\'{a}ria \'{e}
tamb\'{e}m uma soma ponderada, mas com os pesos diminuindo para tr\'{a}s no
tempo \cite{Podlubny, Podlubny1, Podlubny2, Samko}.

Em termos matem\'{a}ticos, o CF atrai por si pr\'{o}prio um grande interesse,
pois o formalismo envolvido faz uso de diversas fun\c{c}\~{o}es especiais,
tais como a generaliza\c{c}\~{a}o da fun\c{c}\~{a}o fatorial, que \'{e} a
fun\c{c}\~{a}o gama e a fun\c{c}\~{a}o de Mittag-Leffler de um par\^{a}metro,
como uma generaliza\c{c}\~{a}o da fun\c{c}\~{a}o exponencial, dentre muitas
outras de interesse acad\^{e}mico e pr\'{a}tico \cite{Mainardi, Gorenflo}.
Existe mais de uma formula\c{c}\~{a}o poss\'{\i}vel para o CF, sendo cada uma
dessas mais adequada a um certo contexto f\'{\i}sico do que outro. As
defini\c{c}\~{o}es mais comuns s\~{a}o as de Riemann-Liouville, Caputo,
Gr\"{u}nwald-Letnikov, Liouville, Weyl e Riesz-Feller \cite{TeseFabio}. Parece
que o n\'{u}mero de defini\c{c}\~{o}es n\~{a}o para de crescer \cite{Capelas}.
S\~{a}o recentes as formula\c{c}\~{o}es de Hilfer \cite{Hilfer}, para
particulares valores do par\^{a}metro associado \`{a} ordem da derivada, onde
as formula\c{c}\~{o}es de Riemann-Liouville e Caputo s\~{a}o casos extremos e
a formula\c{c}\~{a}o proposta por Khalil-Horani-Yousef-Sababheh \cite{Khalil}
a assim chamada \emph{conformable fractional derivative} que, numa conveniente
tradu\c{c}\~{a}o para o portugu\^{e}s, pode ser chamada de derivada
fracion\'{a}ria compat\'{\i}vel.

Neste trabalho discutem-se modelos fracion\'{a}rios associados \`{a} derivada
de Riemann-Liouville e de Caputo, no sentido de que tais modelos s\~{a}o mais
representativos que os respectivos modelos lineares cl\'{a}ssicos de ordem
inteira, isto \'{e}, descrevem com maior acur\'{a}cia os sistemas em quest\~{a}o.

O trabalho est\'{a} disposto da seguinte forma: Na se\c{c}\~{a}o dois s\~{a}o
introduzidos os operadores de integra\c{c}\~{a}o e os operadores de
diferencia\c{c}\~{a}o fracion\'{a}rios conforme propostos por
Riemann-Liouville e Caputo. Na terceira se\c{c}\~{a}o s\~{a}o abordadas as
equa\c{c}\~{o}es diferenciais fracion\'{a}rias (EDF) conforme as
formula\c{c}\~{o}es supra-citadas, sendo que para estas duas
formula\c{c}\~{o}es apresentam-se, atrav\'{e}s de teoremas, os casos gerais.
Na se\c{c}\~{a}o quatro, discute-se a transformada de Laplace associada \`{a}s
integrais e derivadas fracion\'{a}rias e justifica-se a conveni\^{e}ncia de se
trabalhar com a formula\c{c}\~{a}o da derivada segundo Caputo. Ainda na
se\c{c}\~{a}o quatro apresenta-se, atrav\'{e}s de exemplos, o c\'{a}lculo da
transformada de Laplace de fun\c{c}\~{o}es de Mittag-Leffler, concluindo a
se\c{c}\~{a}o, resolve-se duas EDFs atrav\'{e}s da transformada de Laplace,
cuja solu\c{c}\~{a}o \'{e} dada em termos de fun\c{c}\~{o}es de Mittag-Leffler
de dois par\^{a}metros. Na se\c{c}\~{a}o cinco discute-se a modelagem de dois
sistemas onde o c\'{a}lculo fracion\'{a}rio desempenha papel preponderante.
Ap\'{o}s uma revis\~{a}o do conceito de viscoelasticidade linear, em
particular, discutindo os cl\'{a}ssicos modelos de Maxwell e Voigt,
apresenta-se o modelo fracion\'{a}rio de Scott-Blair o qual, nos limites
extremos, recupera os dois modelos cl\'{a}ssicos. O modelo fracion\'{a}rio tem
sua justificativa atrav\'{e}s de gr\'{a}ficos a partir dos quais fica clara a
import\^{a}ncia da formula\c{c}\~{a}o. Por fim, o cl\'{a}ssico problema do
oscilador harm\^{o}nico em sua vers\~{a}o fracion\'{a}ria \'{e} discutido
sendo a solu\c{c}\~{a}o dada em termos de fun\c{c}\~{o}es de Mittag-Leffler
com um e dois par\^{a}me-tros \cite{Gorenflo}. Aqui tamb\'{e}m, o caso limite
(cl\'{a}ssico) \'{e} recuperado no caso em que o par\^{a}metro associado \`{a}
derivada \'{e} igual a dois.

\section{Operadores de Integra\c{c}\~{a}o e Diferencia\c{c}\~{a}o}

Visto que a defini\c{c}\~{a}o de derivada de ordem n\~{a}o inteira nas
formula\c{c}\~{o}es de Riemann-Liouville e Caputo dependem da integral
fracion\'{a}ria, come\c{c}amos com tal conceito, isto \'{e}, introduzimos a
integral fracion\'{a}ria para depois apresentar a derivada fracion\'{a}ria.

Dentre as diversas formula\c{c}\~{o}es existentes para os operadores de
integra\c{c}\~{a}o e diferencia-\c{c}\~{a}o fracion\'{a}rias \cite{Tenreiro1,
Capelas, Fabio}, abordamos, para os prop\'{o}sitos deste trabalho, apenas as
vers\~{o}es segundo Riemann-Liouville e segundo Caputo, definidas a seguir.

\begin{definition}
\label{IFRL}Sejam $\Omega=\left[  a,b\right]  \subset%
\mathbb{R}
$ um intervalo finito e $f\in L_{1}\left[  a,b\right]  $. As express\~{o}es
$\mathcal{I}_{a+}^{\nu}f$ e $\mathcal{I}_{b-}^{\nu}f$ definidas por:%
\begin{equation}
\left(  \mathcal{I}_{a+}^{\nu}f\right)  \left(  x\right)  \equiv\frac
{1}{\Gamma\left(  \nu\right)  }%
{\displaystyle\int\nolimits_{a}^{x}}
\left(  x-t\right)  ^{\nu-1}f\left(  t\right)  \mathbf{d}t,\label{RL-Inta}%
\end{equation}
com $x>a$, $\nu>0$ e%
\begin{equation}
\left(  \mathcal{I}_{b-}^{\nu}f\right)  \left(  x\right)  \equiv\frac
{1}{\Gamma\left(  \nu\right)  }%
{\displaystyle\int\nolimits_{x}^{b}}
\left(  t-x\right)  ^{\nu-1}f\left(  t\right)  \mathbf{d}t,\label{RL-Intb}%
\end{equation}
com $x<b$, $\nu>0$, onde $\Gamma\left(  \nu\right)  $ \'{e} a fun\c{c}\~{a}o
gama, definem as integrais fracion\'{a}rias de
Riemann-Liouville\footnote{Alguns autores \cite{Herrmann, Miller}, distinguem
a nomenclatura, baseando-se nos limites inferior e superior das integrais.
Segundo suas nomenclaturas, Eq.(\ref{RL-Inta}) e Eq.(\ref{RL-Intb}) s\~{a}o
chamadas de vers\~{a}o segundo Riemann; e quando $a=-\infty$ e $b=\infty$,
denotam por vers\~{a}o segundo Liouville.} \emph{(IFRL)} de ordem $\nu\in%
\mathbb{R}
$ num intervalo real finito. As integrais em \emph{Eq.(\ref{RL-Inta})} e
\emph{Eq.(\ref{RL-Intb})} s\~{a}o chamadas de integrais fracion\'{a}rias \`{a}
esquerda e \`{a} direita, respectivamente.
\end{definition}

\begin{definition}
\label{DFRL}Sejam $\Omega=\left[  a,b\right]  \subset%
\mathbb{R}
$ um intervalo finito e $f\in AC^{\mathbf{n}}\left[  a,b\right]  $. As
express\~{o}es $\mathcal{D}_{a+}^{\nu}f$ e $\mathcal{D}_{b-}^{\nu}f$ definidas
por%
\begin{align}
\left(  \mathcal{D}_{a+}^{\nu}f\right)  \left(  x\right)   &  \equiv
\mathcal{D}_{a+}^{\mathbf{n}}\left[  \left(  \mathcal{I}_{a+}^{\mathbf{n}-\nu
}f\right)  \left(  x\right)  \right] \nonumber\\
&  =\left(  \frac{\mathbf{d}}{\mathbf{d}x}\right)  ^{\mathbf{n}}\left(
\mathcal{I}_{a+}^{\mathbf{n}-\nu}f\right)  \left(  x\right) \nonumber\\
&  =\frac{\left(  \frac{\mathbf{d}}{\mathbf{d}x}\right)  ^{\mathbf{n}}}%
{\Gamma\left(  \mathbf{n}-\nu\right)  }%
{\displaystyle\int\nolimits_{a}^{x}}
\frac{f\left(  t\right)  \mathbf{d}t}{\left(  x-t\right)  ^{1+\nu-\mathbf{n}}%
},\label{RL-Dera}%
\end{align}
com $x>a$ e%
\begin{align}
\left(  \mathcal{D}_{b-}^{\nu}f\right)  \left(  x\right)   &  \equiv
\mathcal{D}_{a+}^{\mathbf{n}}\left[  \left(  \mathcal{I}_{b-}^{\mathbf{n}-\nu
}f\right)  \left(  x\right)  \right] \nonumber\\
&  =\left(  -\frac{\mathbf{d}}{\mathbf{d}x}\right)  ^{\mathbf{n}}\left(
\mathcal{I}_{b-}^{\mathbf{n}-\nu}f\right)  \left(  x\right) \nonumber\\
&  =\frac{\left(  -\frac{\mathbf{d}}{\mathbf{d}x}\right)  ^{\mathbf{n}}%
}{\Gamma\left(  \mathbf{n}-\nu\right)  }%
{\displaystyle\int\nolimits_{x}^{b}}
\frac{f\left(  t\right)  \mathbf{d}t}{\left(  t-x\right)  ^{1+\nu-\mathbf{n}}%
}\text{,}\label{RL-Derb}%
\end{align}
com $x<b$, onde $\mathbf{n}=\left[  \nu\right]  +1$ e $\left[  \nu\right]  $
\'{e} a parte inteira de $\nu$, definem as derivadas fracion\'{a}rias de
Riemann-Liouville \emph{(DFRL)} de ordem $\nu\in%
\mathbb{R}
$, com $\nu\geq0$, \`{a} esquerda e \`{a} direita, respectivamente.
\end{definition}

\begin{definition}
\label{DFC}Sejam $\Omega=\left[  a,b\right]  \subset%
\mathbb{R}
$ e $\nu\in%
\mathbb{R}
$, com $\nu>0$. Considere $_{a}\mathcal{D}_{x}^{\mathbf{\nu}}$ e
$_{x}\mathcal{D}_{b}^{\mathbf{\nu}}$ as \emph{DFRL} como nas
\emph{Eq.(\ref{RL-Dera})} e \emph{Eq.(\ref{RL-Derb})} e defina $\mathbf{n}%
=\left[  \nu\right]  +1$, $\nu\notin%
\mathbb{N}
_{0}$; $\mathbf{n}=\nu$ se $\nu\in%
\mathbb{N}
_{0}$, ent\~{a}o as express\~{o}es%
\begin{equation}
_{a}^{C}\mathcal{D}_{x}^{\nu}f(x)=_{a}\mathcal{D}_{x}^{\mathbf{\nu}}\left[
f(x)-%
{\displaystyle\sum\limits_{k=0}^{\mathbf{n}-1}}
\frac{f^{(k)}(a)}{k!}(x-a)^{k}\right] \label{DFCaa}%
\end{equation}
e
\begin{equation}
_{x}^{C}\mathcal{D}_{b}^{\nu}f(x)=_{x}\mathcal{D}_{b}^{\mathbf{\nu}}\left[
f(x)-%
{\displaystyle\sum\limits_{k=0}^{\mathbf{n}-1}}
\frac{f^{(k)}(a)}{k!}(b-x)^{k}\right] \label{DFCbb}%
\end{equation}
definem as derivadas fracion\'{a}rias de Caputo \`{a} esquerda e \`{a}
direita, respectivamente.
\end{definition}

Observamos inicialmente, que todas as defini\c{c}\~{o}es acima se reduzem aos
casos cl\'{a}ssicos quando escolhemos um valor inteiro para a ordem $\nu$.
Al\'{e}m disso, as defini\c{c}\~{o}es que apresentamos foram elaboradas para
um $\nu\in%
\mathbb{R}
$, no entanto, elas s\~{a}o igualmente v\'{a}lidas se considerarmos $\nu\in%
\mathbb{C}
$, com $\operatorname{Re}\left(  \nu\right)  \geq0$, \cite{Kilbas, Samko}.

\section{Equa\c{c}\~{o}es Diferenciais Fracion\'{a}rias\label{EDFO}}

Nesta se\c{c}\~{a}o apresentamos as chamadas EDF e alguns resultados sobre a
exist\^{e}ncia e unicidade das solu\c{c}\~{o}es, quando definidas, em
intervalos finitos. Observamos que existem v\'{a}rios estudos com resultados
eventualmente distintos para o problema de exist\^{e}ncia e unicidade de
solu\c{c}\~{o}es destas, isto porque cada formula\c{c}\~{a}o depende das
defini\c{c}\~{o}es usadas para os operadores de diferencia\c{c}\~{a}o e
integra\c{c}\~{a}o \cite{Kilbas, Miller, Podlubny, Samko}. Os problemas mais
estudados envolvem as solu\c{c}\~{o}es de EDFs segundo Riemann-Liouville e
segundo Caputo: o primeiro por ser a defini\c{c}\~{a}o mais difundida e o
segundo pela conhecida interpreta\c{c}\~{a}o f\'{\i}sica das condi\c{c}\~{o}es
iniciais (ou de fronteira).

\subsection{Formula\c{c}\~{a}o Segundo Riemann-Liouville}

Uma equa\c{c}\~{a}o diferencial (n\~{a}o linear) fracion\'{a}ria de ordem
$\nu>0$, definida num intervalo finito $\left[  a,b\right]  $ \'{e} da forma%
\begin{equation}
\left(  \mathcal{D}_{a+}^{\nu}y\right)  \left(  x\right)  =f\left(
x,y(x)\right)  ,\label{EDF}%
\end{equation}
onde $\mathcal{D}_{a+}^{\nu}$ \'{e} o operador de DFRL.

Se estabelecermos um conjunto de condi\c{c}\~{o}es iniciais (ou de fronteira)%
\begin{align}
\left(  \mathcal{D}_{a+}^{\nu-k-1}y\right)  \left(  a^{+}\right)   &
=\lim_{x\rightarrow a^{+}}\left(  \mathcal{D}_{a+}^{\nu-k-1}y\right)  \left(
x\right) \label{Cond-ini}\\
&  =b_{k},\text{ }k=0,1,\ldots,\mathbf{n}-1,\nonumber
\end{align}
onde $b_{k}\in%
\mathbb{R}
$, $\mathbf{n}=\left[  \nu\right]  +1$ se $\nu\notin%
\mathbb{N}
$ e $\nu=n$ se $\nu\in%
\mathbb{N}
$, ent\~{a}o analogamente ao caso de ordem inteira\footnote{Obtido quando
$\nu\in%
\mathbb{N}
$.}, denotamos a Eq.(\ref{EDF}) junto com as condi\c{c}\~{o}es
Eq.(\ref{Cond-ini}) de um problema de Cauchy (fracion\'{a}rio). Chamamos
aten\c{c}\~{a}o para o caso em que $k=\mathbf{n}-1$ nas condi\c{c}\~{o}es
acima, pois devemos interpretar%
\begin{align*}
\left(  \mathcal{D}_{a+}^{\nu-\mathbf{n}}y\right)  \left(  a^{+}\right)   &
=\lim_{x\rightarrow a^{+}}\left(  \mathcal{D}_{a+}^{\nu-\mathbf{n}}y\right)
\left(  x\right) \\
&  =\lim_{x\rightarrow a^{+}}\left(  \mathcal{I}_{a+}^{\mathbf{n}-\nu
}y\right)  \left(  x\right)  .
\end{align*}

Foge do escopo deste trabalho a discuss\~{a}o da exist\^{e}ncia e unicidade da
solu\c{c}\~{a}o do problema de Cauchy fracion\'{a}rio, entretanto, mencionamos
que tais resultados j\'{a} se encontram formulados na literatura e, podem ser
encontrados em \cite{Kilbas, Samko, Kai}, por exemplo. Ressaltamos tamb\'{e}m
que o problema de Cauchy (vide Eq.(\ref{EDF}) e Eq.(\ref{Cond-ini})) pode ser
formulado, equivalentemente, em termos de uma equa\c{c}\~{a}o integral de
Volterra \cite{Kilbas, Samko, Kai}.

Para os prop\'{o}sitos deste trabalho, entretanto, a formula\c{c}\~{a}o do
problema de Cauchy como explicitado acima \'{e} muito geral, assim nos
restringimos a discutir exemplos para o caso em que as respectivas
equa\c{c}\~{o}es diferenciais ordin\'{a}rias lineares fracion\'{a}rias (EDOLF)
s\~{a}o da forma%
\begin{equation}
\mathcal{D}_{0+}^{\nu}y(t)-\lambda y(t)=f(t)\label{LFDE}%
\end{equation}
com $\lambda\in%
\mathbb{R}
$ e as condi\c{c}\~{o}es iniciais%
\begin{equation}
\left[  \mathcal{D}_{0+}^{\nu-k-1}y(t)\right]  _{t=0}=b_{k},\text{
}k=0,1,\ldots,\mathbf{n}-1\text{,}\label{Condini}%
\end{equation}
onde $\mathbf{n}=\left[  \nu\right]  +1$. Observamos ainda que o limite
inicial de integra\c{c}\~{a}o foi escolhido como sendo o ponto $t=0$, pois
mesmo que tenhamos inicialmente um outro ponto $t=a$, sempre \'{e}
poss\'{\i}vel por meio de uma mudan\c{c}a de vari\'{a}vel fazer uma
transla\c{c}\~{a}o para a origem.

A solu\c{c}\~{a}o deste problema que, em geral, \'{e} conduzido a uma integral
de Volterra apresenta solu\c{c}\~{a}o em termos de uma integral envolvendo uma
fun\c{c}\~{a}o de Mittag-Leffler com dois par\^{a}me-tros $E_{\alpha,\beta
}\left(  \cdot\right)  $, a qual explicitamos a seguir, para refer\^{e}ncia futura,%

\begin{align}
y(t)  &  =\sum_{k=0}^{\mathbf{n}-1}b_{k}t^{\nu-k-1}E_{\nu,\nu-k}\left(
\lambda t^{\nu}\right) \nonumber\\
&  +\int_{0}^{t}\frac{E_{\nu,\nu}\left(  \lambda\left(  t-\tau\right)  ^{\nu
}\right)  f\left(  \tau\right)  }{\left(  t-\tau\right)  ^{1-\nu}}%
\mathbf{d}\tau,\label{SolCauchy}%
\end{align}
dando-nos uma solu\c{c}\~{a}o expl\'{\i}cita para a equa\c{c}\~{a}o integral
de Volterra associada ao problema de Cauchy composto pelas Eq.(\ref{LFDE}) e
Eq.(\ref{Condini}) \cite{Kilbas, TeseFabio, Kai}.

\subsection{Formula\c{c}\~{a}o Segundo\ Caputo}

Analogamente \`{a} formula\c{c}\~{a}o segundo Riemann-Liouville, consideramos
uma equa\c{c}\~{a}o diferencial (n\~{a}o linear) fracion\'{a}ria de ordem
$\nu>0$, definida num intervalo finito $\left[  a,b\right]  $ da forma%
\begin{equation}
\left(  _{a}^{C}\mathcal{D}_{x}^{\nu}y\right)  (x)=f\left(  x,y(x)\right)
,\label{EDFC}%
\end{equation}
onde $_{a}^{C}\mathcal{D}_{x}^{\nu}$ \'{e} o operador de DFC, sujeita \`{a}s
condi\c{c}\~{o}es iniciais%
\begin{equation}
\left(  _{a}^{C}\mathcal{D}_{x}^{k}y\right)  \left(  0\right)  =b_{k}\text{,
}k=0,1,\ldots,\mathbf{n}-1,\label{Cond-iniC}%
\end{equation}
onde $\mathbf{n}=\left[  \nu\right]  +1$ se $\nu\notin%
\mathbb{N}
$ e $\nu=n$ se $\nu\in%
\mathbb{N}
$. Observamos que na formula\c{c}\~{a}o do problema de Cauchy segundo Caputo,
as derivadas
\[
_{a}^{C}\mathcal{D}_{x}^{k}=\frac{\mathbf{d}^{k}}{\mathbf{d}x^{k}}%
\]
s\~{a}o as pr\'{o}prias derivadas usuais de ordem inteira, o que nos leva
\`{a} tradicional interpreta\c{c}\~{a}o f\'{\i}sica das condi\c{c}\~{o}es
iniciais, como no usual problema de Cauchy para ordens inteiras.

O teorema de exist\^{e}ncia para o problema de Cauchy pode ser encontrado em
\cite{Kai} e, equiva-lentemente \`{a} formula\c{c}\~{a}o de Riemann-Liouville,
este problema pode ser conduzido a uma integral de Volterra (n\~{a}o-linear)
\cite{TeseFabio, Kai}.

Nessa formula\c{c}\~{a}o, o teorema de exist\^{e}ncia para o problema de
Cauchy supracitado fica ent\~{a}o enunciado da seguinte forma \cite{Kai}.

Ressalta-se que os resultados entre as formula\c{c}\~{o}es de
Riemann-Liouville e de Caputo, s\~{a}o tais que enquanto na primeira garante
uma solu\c{c}\~{a}o cont\'{\i}nua somente em $\left(  0,h\right]  $ na segunda
a solu\c{c}\~{a}o \'{e} garantida ser cont\'{\i}nua em $\left[  0,h\right]  $
\cite{TeseFabio, Kai}.

Novamente, para os prop\'{o}sitos deste trabalho as formula\c{c}\~{o}es acima
s\~{a}o muito gerais, assim nos restringimos a discutir exemplos para o caso
das EDOLF da forma%
\begin{equation}
_{0}^{C}\mathcal{D}_{x}^{\nu}y(t)-\lambda y(t)=f(t),\label{LFDEC}%
\end{equation}
com $\lambda\in%
\mathbb{R}
$, junto com as condi\c{c}\~{o}es iniciais%
\begin{equation}
\left[  _{0}^{C}\mathcal{D}_{x}^{k}y(t)\right]  _{t=0}=b_{k}\text{,
}k=0,1,\ldots,\mathbf{n}-1.\label{CondininC}%
\end{equation}

E, novamente, uma vez conduzida a sua forma integral, \'{e} poss\'{\i}vel
obter uma solu\c{c}\~{a}o expl\'{\i}cita \cite{Kilbas} para o problema de
Cauchy composto pelas equa\c{c}\~{o}es Eq.(\ref{LFDEC}) e Eq.(\ref{CondininC})
em termos das fun\c{c}\~{o}es de Mittag-Leffler como%
\begin{align}
y(t)  &  =\sum_{k=0}^{\mathbf{n}-1}b_{k}t^{\nu}E_{\nu,k+1}\left(  \lambda
t^{\nu}\right)  +\nonumber\\
&  \int_{0}^{t}\frac{E_{\nu,\nu}\left(  \lambda\left(  t-\tau\right)  ^{\nu
}\right)  f\left(  \tau\right)  }{\left(  t-\tau\right)  ^{1-\nu}}%
\mathbf{d}\tau.\label{SolCauchyC}%
\end{align}

\section{Transformada de Laplace dos Operadores Fracion\'{a}rios}

Sabemos do c\'{a}lculo de ordem inteira que o m\'{e}todo da transformada de
Laplace \'{e} uma ferramenta muito \'{u}til na an\'{a}lise de equa\c{c}\~{o}es
dife-renciais lineares, principalmente no caso em que a equa\c{c}\~{a}o possui
coeficientes constantes. Neste caso, a transformada de Laplace reduz a EDO
numa equa\c{c}\~{a}o alg\'{e}brica que \'{e}, em geral, muito mais simples de
se solucionar, deixando a dificuldade de se obter a solu\c{c}\~{a}o final da
EDO de partida a um problema de invers\~{a}o. Nesta se\c{c}\~{a}o, veremos que
a mesma metodologia tamb\'{e}m pode ser usada para resolver problemas de valor
inicial relacionado com as EDOLF (vide Eq.(\ref{LFDE}) e Eq.(\ref{LFDEC})).

Lembremos que uma fun\c{c}\~{a}o $f\left(  t\right)  $ \'{e} dita de ordem
exponencial $\alpha$ se existir constantes positivas $M$ e $T$, tais que%
\[
e^{-\alpha t}\left\vert f(t)\right\vert \leq M,
\]
para todo $t\geq T$. Dessa forma, dada uma $f:\left[  0,\infty\right)
\rightarrow%
\mathbb{R}
$, de ordem exponencial $\alpha$ a fun\c{c}\~{a}o $F$ definida por%
\begin{equation}
F(s)=\int_{0}^{\infty}e^{-st}f(t)\mathbf{d}t,\label{TLaplace}%
\end{equation}
\'{e} chamada a transformada de Laplace da $f$, com a condi\c{c}\~{a}o
$\operatorname{Re}\left(  s\right)  >\alpha$ que garante a exist\^{e}ncia da
integral acima. Da nomenclatura cl\'{a}ssica \'{e} usual denotar a
transformada de Laplace de $f$ por $%
\mathcal{L}%
\left[  f(t)\right]  =F(s)$ e iremos denotar por $%
\mathcal{L}%
^{-1}\left[  F(s)\right]  =f(t)$ a transformada de Laplace inversa, que como
sabemos \'{e} \'{u}nica pelo teorema de Lerch \cite{Capelas3}.

Dentre as diversas propriedades conhecidas da transformada de Laplace,
listamos as seguintes:

\begin{itemize}
\item \textbf{Linearidade: }$\mathcal{%
\mathcal{L}%
}\left[  \alpha f(t)+\beta g(t)\right]  =\alpha%
\mathcal{L}%
\left[  f(t)\right]  +\beta%
\mathcal{L}%
\left[  g(t)\right]  =\alpha F(s)+\beta G(s)$, com $\alpha,\beta\in%
\mathbb{C}
$;

\item \textbf{Convolu\c{c}\~{a}o: }$%
\mathcal{L}%
\left[  f\left(  t\right)  \star g(t)\right]  =%
\mathcal{L}%
\left[  f(t)\right]
\mathcal{L}%
\left[  g(t)\right]  =F(s)G(s)$, onde o produto $\star$ \'{e} o de convolu\c{c}\~{a}o;

\item \textbf{Derivadas: }%
\[%
\mathcal{L}%
\left[  \frac{\mathbf{d}^{n}f}{\mathbf{d}t^{n}}\right]  =s^{n}F(s)-\sum
\limits_{k=0}^{n-1}s^{n-k-1}f^{(k)}(0^{+});
\]

\item \textbf{Integral: }%
\[%
\mathcal{L}%
\left[  \int_{0}^{t}f\left(  \tau\right)  \mathbf{d}\tau\right]  =\frac{1}%
{s}F(s).
\]

\end{itemize}

Assim, nosso intuito \'{e} calcular as transformadas de Laplace da IFRL, da
DFRL e da DFC de uma dada fun\c{c}\~{a}o $f\left(  t\right)  $ suficientemente
bem comportada\footnote{No sentido de satisfazer as condi\c{c}\~{o}es de
exist\^{e}ncia das IFRL/DFRL e das respectivas transformadas.}.

Lembrando que a IFRL pode ser escrita como o produto de convolu\c{c}\~{a}o
\cite{Fabio}:%
\begin{align*}
\left(  \mathcal{I}_{a+}^{\nu}f\right)  \left(  t\right)   &  \equiv\frac
{1}{\Gamma\left(  \nu\right)  }%
{\displaystyle\int\nolimits_{a}^{t}}
\left(  t-\tau\right)  ^{\nu-1}f\left(  \tau\right)  \mathbf{d}\tau\\
&  =f\left(  t\right)  \star\phi_{\nu}\left(  t\right)  ,
\end{align*}
ent\~{a}o segue da transformada do produto de convolu\c{c}\~{a}o listada
acima, que%
\begin{align}%
\mathcal{L}%
\left[  \left(  \mathcal{I}_{0+}^{\nu}f\right)  \left(  t\right)  \right]   &
=%
\mathcal{L}%
\left[  f\left(  t\right)  \star\phi_{\nu}\left(  t\right)  \right]
\label{LaplaceIFRL}\\
&  =F(s)\frac{1}{s^{\nu}}.\nonumber
\end{align}

Agora da \textbf{Defini\c{c}\~{a}o \ref{DFRL}}, temos que $\left(
\mathcal{D}_{0+}^{\nu}f\right)  \left(  t\right)  \equiv\mathcal{D}%
_{0+}^{\mathbf{n}}g(t) $, com $\mathbf{n=}\left[  \nu\right]  +1$ onde
$g(t)=\left(  \mathcal{I}_{0+}^{\mathbf{n}-\nu}f\right)  \left(  t\right)  $,
assim%
\begin{align}%
\mathcal{L}%
\left[  \left(  \mathcal{D}_{0+}^{\nu}f\right)  \left(  t\right)  \right]   &
=%
\mathcal{L}%
\left[  \mathcal{D}_{0+}^{\mathbf{n}}g(t)\right] \label{LDFRLa}\\
&  =s^{\mathbf{n}}G(s)-\sum_{k=0}^{\mathbf{n}-1}s^{\mathbf{n}-k-1}g^{\left(
k\right)  }(0^{+}).\nonumber
\end{align}
Mas, por outro lado, temos%
\begin{align}
G(s)  &  =%
\mathcal{L}%
\left[  \left(  \mathcal{I}_{0+}^{\mathbf{n}-\nu}f\right)  \left(  t\right)
\right] \nonumber\\
&  =\frac{F(s)}{s^{\mathbf{n}-\nu}},\label{LDFRLb}%
\end{align}
portanto, usando a Eq.(\ref{LDFRLb}) na Eq.(\ref{LDFRLa}), obtemos%
\begin{equation}%
\mathcal{L}%
\left[  \left(  \mathcal{D}_{0+}^{\nu}f\right)  \left(  t\right)  \right]
=s^{\nu}F(s)-\sum_{k=0}^{\mathbf{n}-1}s^{\mathbf{n}-k-1}g^{\left(  k\right)
}(0^{+}),\label{LaplaceDFRL}%
\end{equation}
onde $g^{\left(  k\right)  }(0^{+})=\lim\limits_{t\rightarrow0^{+}}%
\mathcal{D}^{k}\left(  \mathcal{I}_{0+}^{\mathbf{n}-\nu}f\right)  \left(
t\right)  $.

Por outro lado, quando usamos a \textbf{Defini\c{c}\~{a}o \ref{DFC}}, podemos
escrever que%
\begin{align}%
\mathcal{L}%
\left[  \left(  _{0}^{C}\mathcal{D}_{x}^{\nu}f\right)  \left(  t\right)
\right]   &  =%
\mathcal{L}%
\left[  \mathcal{I}_{0+}^{\mathbf{n}-\nu}g(t)\right] \nonumber\\
&  =\frac{1}{s^{\mathbf{n}-\nu}}G(s),\label{LDFRLaC}%
\end{align}
onde%
\begin{align}
G(s)  &  =%
\mathcal{L}%
\left[  \left(  \mathcal{D}_{0+}^{\mathbf{n}}f\right)  \left(  t\right)
\right] \nonumber\\
&  =s^{\mathbf{n}}F(s)-\sum_{k=0}^{\mathbf{n}-1}s^{\mathbf{n}-k-1}f^{\left(
k\right)  }(0^{+}).\label{LDFRLbC}%
\end{align}

Logo, os resultados das Eq.(\ref{LDFRLaC}) e Eq.(\ref{LDFRLbC}) nos levam
\`{a} seguinte express\~{a}o%
\begin{equation}%
\mathcal{L}%
\left[  \left(  _{0}^{C}\mathcal{D}_{x}^{\nu}f\right)  \left(  t\right)
\right]  =s^{\nu}F(s)-\sum_{k=0}^{\mathbf{n}-1}s^{\mathbf{\nu}-k-1}f^{\left(
k\right)  }(0^{+}).\label{LaplaceDFC}%
\end{equation}

Novamente fica evidente a diferen\c{c}a entre as defini\c{c}\~{o}es dos
operadores de DFRL e DFC, sendo que a segunda se mostra nitidamente mais
oportuna de ser usada quando temos interpreta\c{c}\~{o}es f\'{\i}sicas claras
das condi\c{c}\~{o}es iniciais do problema de Cauchy, como nos mostram os
resultados das transformadas de Laplace para a DFRL e DFC (vide
Eq.(\ref{LaplaceDFRL}) e Eq.(\ref{LaplaceDFC})).

Antes de apresentarmos alguns exemplos, listamos, por conveni\^{e}ncia, as
transformadas de Laplace das fun\c{c}\~{o}es de Mittag-Leffler\footnote{Estas
transformadas podem ser facilmente calculadas, usando a representa\c{c}\~{a}o
em s\'{e}rie das fun\c{c}\~{o}es de Mittag-Leffler e calculando as
transformadas termo a termo, o que \'{e} poss\'{\i}vel visto que estas
fun\c{c}\~{o}es s\~{a}o inteiras no plano $%
\mathbb{C}
$.}, que nos ser\~{a}o \'{u}teis na hora de resolvermos o problema de
invers\~{a}o $%
\mathcal{L}%
^{-1}:F(s)\mapsto f(t)$. Assim, temos%
\begin{align}%
\mathcal{L}%
\left[  E_{\alpha}(-\lambda t^{\alpha})\right]   &  =\frac{s^{\alpha-1}%
}{s^{\alpha}+1}\nonumber\\
&  =\frac{s^{-1}}{1+\lambda s^{-\alpha}}\label{LE1}\\%
\mathcal{L}%
\left[  t^{\beta-1}E_{\alpha,\beta}(-\lambda t^{\alpha})\right]   &
=\frac{s^{\alpha-\beta}}{s^{\alpha}+\lambda}\nonumber\\
&  =\frac{s^{-\beta}}{1+\lambda s^{-\alpha}}\label{LE2}\\%
\mathcal{L}%
\left[  t^{\beta-1}E_{\alpha,\beta}^{\gamma}(-\lambda t^{\alpha})\right]   &
=\frac{s^{\alpha\gamma-\beta}}{\left(  s^{\alpha}+\lambda\right)  ^{\gamma}%
}\nonumber\\
&  =\frac{s^{-\beta}}{\left(  1+\lambda s^{-\alpha}\right)  ^{\gamma}%
}\label{LE3}%
\end{align}

Antes de passarmos a discutir modelos via CF, apresentamos dois problemas de
valor inicial envolvendo a formula\c{c}\~{a}o de Riemann-Liouville (um
problema homog\^{e}neo e o outro n\~{a}o homog\^{e}neo) enquanto a
formula\c{c}\~{a}o de Caputo ser\'{a} aprensentado na pr\'{o}xima se\c{c}\~{a}o.

\begin{example}
Determine $y(t)$ no problema%
\[
\left\{
\begin{array}
[c]{l}%
\mathcal{D}_{0+}^{\frac{1}{2}}y(t)-\lambda y(t)=0,\\
\mathcal{D}_{0+}^{\frac{1}{2}-1}y(0^{+})=\mathcal{I}_{0+}^{\frac{1}{2}}%
y(0^{+})=b_{0}\in%
\mathbb{R}
.
\end{array}
\right.
\]
Aplicando a transformada de Laplace \`{a} equa\c{c}\~{a}o, obtemos%
\begin{align*}%
\mathcal{L}%
\left[  \mathcal{D}_{0+}^{\frac{1}{2}}y(t)-\lambda y(t)\right]   &  =0\\
s^{\frac{1}{2}}Y(s)-b_{0}-\lambda Y(s)  &  =0\\
\left(  s^{\frac{1}{2}}-\lambda\right)  Y(s)  &  =b_{0},
\end{align*}
o que implica que em%
\[
Y(s)=\frac{b_{0}}{s^{\frac{1}{2}}-\lambda},
\]
donde podemos verificar, usando a \emph{Eq.(\ref{LE2})}, que%
\[
y(t)=b_{0}t^{-\frac{1}{2}}E_{\frac{1}{2},\frac{1}{2}}(\lambda t^{\frac{1}{2}%
}),
\]
onde $E_{\alpha,\beta}\left(  \cdot\right)  $ \'{e} a fun\c{c}\~{a}o de
Mittag-Leffler com dois par\^{a}metros \emph{\cite{Gorenflo}}. A
solu\c{c}\~{a}o coincide com a solu\c{c}\~{a}o do problema de Cauchy conforme
a \emph{Eq.(\ref{SolCauchy})}.
\end{example}

Vejamos ainda um segundo exemplo, agora n\~{a}o-homog\^{e}neo.

\begin{example}
Resolva o problema composto pela \emph{EDF} e as condi\c{c}\~{o}es a seguir:%
\[%
\begin{array}
[c]{l}%
\mathcal{D}_{0+}^{\frac{4}{3}}y(t)-\lambda y(t)=t^{2},\\
\mathcal{D}_{0+}^{\frac{4}{3}-1}y(0^{+})=\mathcal{D}_{0+}^{\frac{1}{3}}%
y(0^{+})=b_{0}\in%
\mathbb{R}
,\\
\mathcal{D}_{0+}^{\frac{4}{3}-2}y(0^{+})=\mathcal{I}_{0+}^{\frac{2}{3}}%
y(0^{+})=b_{1}\in%
\mathbb{R}
.
\end{array}
\]
\newline Aplicando a transformada de Laplace na equa\c{c}\~{a}o, obtemos%
\[%
\mathcal{L}%
\left[  \mathcal{D}_{0+}^{\frac{4}{3}}y(t)-\lambda y(t)\right]  =\frac
{\Gamma\left(  3\right)  }{s^{3}},
\]
ou ainda%
\[
\left(  s^{\frac{4}{3}}-\lambda\right)  Y(s)=\frac{\Gamma\left(  3\right)
}{s^{3}}+sb_{1}+b_{0},
\]
o que implica em%
\[
Y(s)=\frac{b_{0}}{s^{\frac{4}{3}}-\lambda}+\frac{sb_{1}}{s^{\frac{4}{3}%
}-\lambda}+\frac{\Gamma\left(  3\right)  }{s^{3}}\frac{1}{s^{\frac{4}{3}%
}-\lambda},
\]
donde podemos verificar, usando a \emph{Eq.(\ref{LE2})} e a propriedade da
transformada do produto de convolu\c{c}\~{a}o, que%
\begin{align*}
y(t)  &  =b_{0}t^{\frac{1}{3}}E_{\frac{4}{3},\frac{4}{3}}\left(  \lambda
t^{\frac{4}{3}}\right)  +b_{1}t^{-\frac{2}{3}}E_{\frac{4}{3},\frac{1}{3}%
}\left(  \lambda t^{\frac{4}{3}}\right) \\
&  +\left[  t^{2}\star\left(  t^{\frac{1}{3}}E_{\frac{4}{3},\frac{4}{3}%
}\left(  \lambda t^{\frac{4}{3}}\right)  \right)  \right] \\
&  =\sum_{k=0}^{1}b_{k}t^{\frac{4}{3}-k-1}E_{\frac{4}{3},\frac{4}{3}-k}\left(
\lambda t^{\frac{4}{3}}\right)  +\\
&  \int_{0}^{t}\left[  \left(  t-\tau\right)  ^{\frac{1}{3}}E_{\frac{4}%
{3},\frac{4}{3}}\left(  \lambda\left(  t-\tau\right)  ^{\frac{4}{3}}\right)
\right]  \tau^{2}\mathbf{d}\tau
\end{align*}
e, novamente, o resultado coincide com a f\'{o}rmula para a solu\c{c}\~{a}o do
problema de Cauchy conforme a \emph{Eq.(\ref{SolCauchy})}.
\end{example}

Estes exemplos apesar de simples, ilustram que a metodologia da transformada
de Laplace continua tendo a mesma efic\'{a}cia para a resolu\c{c}\~{a}o de
EDOLFs (com coeficientes constantes) para a formula\c{c}\~{a}o de
Riemann-Liouville, assim como nos casos de ordens inteiras. Conforme
mencionamos,uma aplica\c{c}\~{a}o usando a formula\c{c}\~{a}o segundo Caputo,
ser\'{a} discutida quando apresentarmos uma poss\'{\i}vel
generaliza\c{c}\~{a}o para o pro-blema de Cauchy associado aos osciladores
harm\^{o}nicos na \textbf{Se\c{c}\~{a}o \ref{OscHarm}}.

\section{Modelos via C\'{a}lculo Fracion\'{a}rio \label{Modelagem}}

Vejamos agora como os conceitos apresentados do c\'{a}lculo fracion\'{a}rio
podem ajudar a descre-ver de forma mais apropriada certos fen\^{o}menos
f\'{\i}sicos, quando comparados com a abordagem cl\'{a}ssica. Ilustraremos a
afirma\c{c}\~{a}o acima investigando um problema particular no campo da
viscoelasticidade linear e numa poss\'{\i}vel genera-liza\c{c}\~{a}o do
problema associado ao oscilador harm\^{o}nico.

\subsection{Viscoelasticidade Linear}

A viscoelasticidade \'{e} uma \'{a}rea que investiga o comportamento de
materiais que admitem carac-ter\'{\i}sticas el\'{a}stica e de viscosidade
quando submetidos \`{a} for\c{c}as de deforma\c{c}\~{a}o. No estudo destas
caracter\'{\i}sticas, dois conceitos s\~{a}o importantes: a tens\~{a}o
(\emph{stress}) $\sigma(t)$ e a deforma\c{c}\~{a}o (\emph{strain})
$\epsilon(t)$, ambas descritas como uma fun\c{c}\~{a}o do tempo $t$ e os
modelos matem\'{a}ticos aplicados procuram, justamente, descrever a
rela\c{c}\~{a}o entre estas quantidades.\footnote{Em geral, a tens\~{a}o e a
deforma\c{c}\~{a}o s\~{a}o campos tensoriais, assim para esta
apresenta\c{c}\~{a}o pressupomos que os materiais s\~{a}o isotropicamente
uniformes.}

Classicamente, as leis da mec\^{a}nica que s\~{a}o usa-das para descrever as
rela\c{c}\~{o}es supracitadas s\~{a}o as de Newton e de Hooke, sendo a
primeira para o comportamento de l\'{\i}quidos ideais (ou l\'{\i}quidos
Newtonianos) modelado pela equa\c{c}\~{a}o%
\begin{equation}
\sigma(t)=\eta\frac{\mathbf{d}}{\mathbf{d}t}\epsilon(t),\label{Newton}%
\end{equation}
onde $\eta$ \'{e} a chamada constante de viscosidade do material. J\'{a} a
segunda lei (Hooke) modelada pela equa\c{c}\~{a}o%
\begin{equation}
\sigma(t)=E\epsilon(t),\label{Hooke}%
\end{equation}
descreve o comportamento de materiais el\'{a}sticos (ideais), sendo $E$ a
chamada constante el\'{a}stica (ou m\'{o}dulo de elasticidade) do material.

No estudo de sistemas modelados pelas equa\c{c}\~{o}es Eq.(\ref{Newton}) ou
Eq.(\ref{Hooke}), o interesse \'{e} obter descri\c{c}\~{o}es sobre as
respostas da tens\~{a}o com res-peito \`{a} deforma\c{c}\~{a}o ou viceversa,
sendo que os ensaios experimentais usualmente feitos s\~{a}o os de
\cite{Mainardi}:

\begin{itemize}
\item \textbf{Relaxa\c{c}\~{a}o de tens\~{a}o via deforma\c{c}\~{a}o
controlada:} onde se mede o estado de tens\~{a}o causado no material de teste
pela aplica\c{c}\~{a}o de uma deforma\c{c}\~{a}o previamente definida. Neste
caso, a resposta da tens\~{a}o \`{a} aplica\c{c}\~{a}o da deforma\c{c}\~{a}o
\'{e} chamada \emph{m\'{o}dulo de relaxa\c{c}\~{a}o} $\mathbf{G}(t)$.

\item \textbf{Teste de flu\^{e}ncia via tens\~{a}o controlada:} onde se
aplicam ao material teste uma tens\~{a}o previamente definida e medem-se as
deforma\c{c}\~{o}es causadas por essa tens\~{a}o aplicada. Neste caso, a
resposta da deforma\c{c}\~{a}o \`{a} aplica\c{c}\~{a}o de uma tens\~{a}o \'{e}
chamada de \emph{flu\^{e}ncia} $\mathbf{J}(t)$.
\end{itemize}

No estudo destes experimentos, as excita\c{c}\~{o}es impostas ao sistema (ou
material teste) s\~{a}o usualmente modeladas pelo impulso (fun\c{c}\~{a}o
ge-neralizada) $\delta$-Dirac ou a fun\c{c}\~{a}o degrau (ou de Heaviside)
$H(t)$ definida por%
\begin{equation}
H(t)=\left\{
\begin{array}
[c]{cc}%
0, & t\leq0,\\
1, & t>0,
\end{array}
\right. \label{Heaviside}%
\end{equation}
sendo que esta \'{u}ltima \'{e} fisicamente mais realista que a primeira.
Logo, neste trabalho, conside-raremos as respostas $\mathbf{G}(t)$ e
$\mathbf{J}(t)$ quando submetidas a uma excita\c{c}\~{a}o do tipo
fun\c{c}\~{a}o degrau $H(t)$ (Ver Figura 1 e Figura 2).%

\begin{figure}[ptb]%
\centering
\includegraphics[
height=1.7945in,
width=2.6628in
]%
{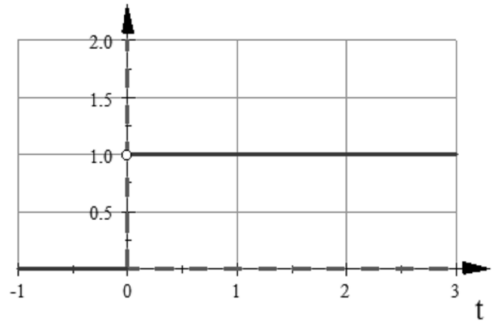}%
\caption{O M\'{o}dulo de relaxa\c{c}\~{a}o para um s\'{o}lido ideal ($E=1$;
tra\c{c}o cont\'{\i}nuo) e um fluido ideal ($\eta=1$; tra\c{c}o pontilhado)
gerados a partir de uma excita\c{c}\~{a}o com a fun\c{c}\~{a}o de Heaviside.}%
\end{figure}
%

\begin{figure}[ptb]%
\centering
\includegraphics[
height=1.7945in,
width=2.6628in
]%
{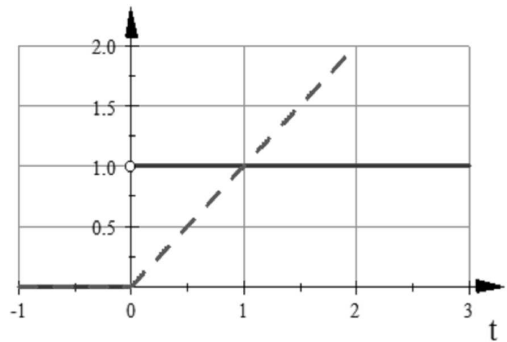}%
\caption{O M\'{o}dulo de flu\^{e}ncia para um s\'{o}lido ideal ($E=1$;
tra\c{c}o cont\'{\i}nuo) e um fluido ideal ($\eta=1$; tra\c{c}o pontilhado)
gerados a partir de uma excita\c{c}\~{a}o com a fun\c{c}\~{a}o de Heaviside.}%
\end{figure}

Na natureza, entretanto, n\~{a}o existem s\'{o}lidos ou fluidos ideais de modo
que na pr\'{a}tica os materiais reais possuem propriedades situadas entre
estes dois casos limites. Esquematicamente, o comportamento el\'{a}stico
(ideal) \'{e} descrito por uma mola, enquanto o comportamento viscoso (ideal)
\'{e} descrito como um amortercedor (Figura 3) e os modelos aplicados na
pr\'{a}tica s\~{a}o uma combina\c{c}\~{a}o destes dois elementos ideais.%

\begin{figure}[ptb]%
\centering
\includegraphics[
height=1.7945in,
width=2.6628in
]%
{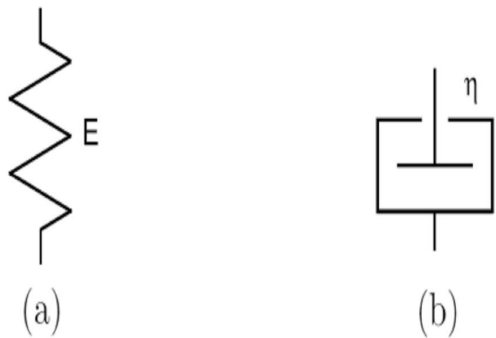}%
\caption{(a) Elemento el\'{a}stico (ideal) \'{e} descrito por uma mola; (b)
enquanto o elemento viscoso (ideal) \'{e} apresentado como um amortecedor.}%
\end{figure}

Entre os modelos mais simples, est\~{a}o os de Maxwell e de Voigt, como
descritos pela Figura 4:%

\begin{figure}[ptb]%
\centering
\includegraphics[
height=1.7945in,
width=2.6628in
]%
{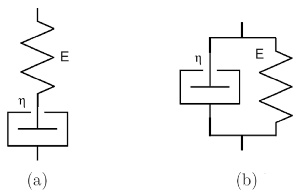}%
\caption{(a) Esquematiza\c{c}\~{a}o em s\'{e}rie (Maxwell); (b)
Esquematiza\c{c}\~{a}o em paralelo (Voigt).}%
\end{figure}

No modelo de Maxwell, a equa\c{c}\~{a}o que des-creve esta
configura\c{c}\~{a}o \'{e} dada por%
\begin{equation}
\frac{1}{\eta}\sigma(t)+\frac{1}{E}\frac{\mathbf{d}}{\mathbf{d}t}%
\sigma(t)=\frac{\mathbf{d}}{\mathbf{d}t}\epsilon(t)\label{Maxwell}%
\end{equation}
e as fun\c{c}\~{o}es de resposta (m\'{o}dulo de relaxa\c{c}\~{a}o
$\mathbf{G}_{M}(t)$ e flu\^{e}ncia $\mathbf{J}_{M}(t)$) s\~{a}o descritas como%
\begin{align}
\mathbf{G}_{M}(t) &  =Ee^{-\frac{E}{\eta}t},\label{MaxG}\\
\mathbf{J}_{M}(t) &  =\frac{1}{E}H(t)+\frac{1}{\eta}t.\label{MaxJ}%
\end{align}

J\'{a} no modelo de Voigt, a equa\c{c}\~{a}o que des-creve esta
configura\c{c}\~{a}o \'{e} dada por%
\begin{equation}
\sigma(t)=E\epsilon(t)+\eta\frac{\mathbf{d}}{\mathbf{d}t}\epsilon
(t)\label{Voigt}%
\end{equation}
e as fun\c{c}\~{o}es de resposta (m\'{o}dulo de relaxa\c{c}\~{a}o
$\mathbf{G}_{V}(t)$ e flu\^{e}ncia $\mathbf{J}_{V}(t)$) s\~{a}o descritas como%
\begin{align}
\mathbf{G}_{V}(t)  &  =EH(t)+\eta\delta(t),\label{VoigtG}\\
\mathbf{J}_{V}(t)  &  =\frac{1}{E}\left(  1-e^{-\frac{E}{\eta}t}\right)
.\label{VoigtJ}%
\end{align}

Estes dois modelos tamb\'{e}m diferem das observa\c{c}\~{o}es experimentais em
alguns pontos de modo que estudiosos da \'{a}rea elaboram os mais diversos
modelos (e.g., Zener, Kelvin, Burger, etc...) para tentar chegar a uma
descri\c{c}\~{a}o com maior sintonia com os experimentos \cite{Mainardi}. Por
exemplo, continuando com a tentativa de ge-neraliza\c{c}\~{a}o dos modelos de
Maxwell e Voigt, os modelos de Zener\footnote{Seguindo a sugest\~{a}o de
nomenclatura como proposto em \cite{Mainardi}, estes modelos foram estudados
por Clarence Melvin Zener (1905-1993) e os resultados publicados no seu livro
de 1948, \emph{Elasticity and Anelasticity of Metals}, Univ. of Chicago Press;
e tamb\'{e}m s\~{a}o conhecidos como modelos de \emph{s\'{o}lidos lineares
padr\~{a}o.}} combinam os dois anteriores dos seguintes modos (Figura 5): (a)
o modelo de Maxwell modificado combina um elemento de mola com o elemento de
Maxwell em paralelo (b) e o modelo de Voigt modificado combina um elemento de
mola em s\'{e}rie com o elemento de Voigt.%

\begin{figure}[ptb]%
\centering
\includegraphics[
height=1.7945in,
width=2.6628in
]%
{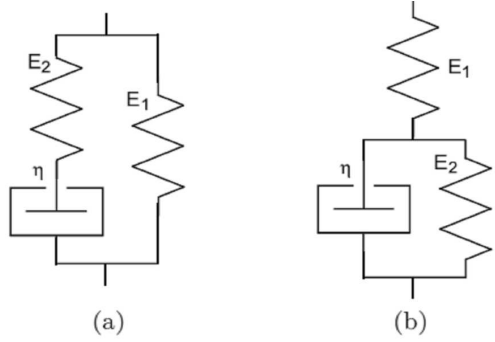}%
\caption{Os dois tipos de modelos estudados por Zener: (a) Maxwell modificado;
(b) Voigt modificado.}%
\end{figure}

O modelo de Maxwell modificado \'{e} descrito pela equa\c{c}\~{a}o%
\begin{equation}
\frac{E_{2}}{\eta}\sigma(t)+\frac{\mathbf{d}\sigma(t)}{\mathbf{d}t}%
=\frac{E_{1}E_{2}}{\eta}\epsilon(t)+\left(  E_{1}+E_{2}\right)  \frac
{\mathbf{d}\epsilon(t)}{\mathbf{d}t},\label{Maxmod}%
\end{equation}
com respostas (m\'{o}dulo de relaxa\c{c}\~{a}o $\mathbf{G}_{ZM}(t)$ e
flu\^{e}ncia $\mathbf{J}_{ZM}(t)$) dadas por, respectivamente%
\begin{align}
\mathbf{G}_{ZM}(t) &  =E_{1}+E_{2}e^{-\frac{E_{2}}{\eta}t},\label{MaxmodG}\\
\mathbf{J}_{ZM}(t) &  =\frac{1}{E_{1}}-\frac{E_{2}e^{-\frac{E_{1}E_{2}}%
{\eta\left(  E_{1}+E_{2}\right)  }t}}{E_{1}\left(  E_{1}+E_{2}\right)
}.\label{MaxmodJ}%
\end{align}

J\'{a} o modelo de Voigt modificado \'{e} descrito pela equa\c{c}\~{a}o%
\begin{equation}
\frac{E_{1}+E_{2}}{\eta}\sigma(t)+\frac{\mathbf{d}\sigma(t)}{\mathbf{d}%
t}=\frac{E_{1}E_{2}}{\eta}\epsilon(t)+E_{1}\frac{\mathbf{d}\epsilon
(t)}{\mathbf{d}t},\label{Voigtmod}%
\end{equation}
com respostas (m\'{o}dulo de relaxa\c{c}\~{a}o $\mathbf{G}_{ZV}(t)$ e
flu\^{e}ncia $\mathbf{J}_{ZV}(t)$) dadas por, respectivamente%
\begin{align}
\mathbf{G}_{ZV}(t) &  =\frac{E_{1}^{2}}{E_{1}+E_{2}}e^{-\frac{E_{1}+E_{2}%
}{\eta}t}+\frac{E_{1}E_{2}}{E_{1}+E_{2}},\label{VoigmodG}\\
\mathbf{J}_{ZV}(t) &  =\frac{E_{1}+E_{2}}{E_{1}E_{2}}-\frac{1}{E_{2}}%
e^{-\frac{E_{2}}{\eta}t}.\label{VoigmodJ}%
\end{align}

A Figura 6, mostra as respostas dos quatro casos mencionados: Maxwell, Voigt,
Maxwell modificado e Voigt Modificado. Observamos que no modelo de Maxwell, o
processo de relaxa\c{c}\~{a}o da tens\~{a}o parece razoavelmente realista, no
entanto, a resposta de flu\^{e}ncia mostra uma deforma\c{c}\~{a}o crescendo
indefinidamente. No mo-delo de Voigt, o contr\'{a}rio acontece, a resposta de
flu\^{e}ncia parece razoavelmente realista, mas o processo de
relaxa\c{c}\~{a}o \'{e} inexistente (permanece constante). Por outro lado, nos
dois modelos modificados as respostas s\~{a}o, at\'{e} certa forma,
semelhantes, sendo que ambos qualitativamente ficam de acordo com alguns
resultados experimentais, mas ambos possuem problemas na des-cri\c{c}\~{a}o
quantitativa, visto que predizem um valor finito para a tens\~{a}o no tempo
inicial $t=0$ mesmo diante do salto (descont\'{\i}nuo) de mudan\c{c}a de
deforma\c{c}\~{a}o causada por uma excita\c{c}\~{a}o do tipo Heaviside $H(t)$
e nenhum dos dois atinge uma relaxa\c{c}\~{a}o de tens\~{a}o completa.
Al\'{e}m disso, ambos modelos modificados predizem uma descontinuidade da
resposta de flu\^{e}ncia imediatamente ap\'{o}s a aplica\c{c}\~{a}o de uma
excita\c{c}\~{a}o de tens\~{a}o do tipo Heaviside $H(t)$. Estudos indicam que
os modelos podem ser aperfei\c{c}oados se forem inclu\'{\i}dos mais elementos
(de Newton, de Hooke, de Maxwell, de Voigt, etc) em s\'{e}rie ou paralelo, mas
isto implicaria em equa\c{c}\~{o}es cada vez mais complicadas e de ordens
superiores. Uma alternativa para encontrarmos uma solu\c{c}\~{a}o mais
satisfat\'{o}ria a este problema, faz uso do c\'{a}lculo fracion\'{a}rio no
equacionamento do problema como veremos a seguir.%

\begin{figure}[ptb]%
\centering
\includegraphics[
height=1.5385in,
width=2.6602in
]%
{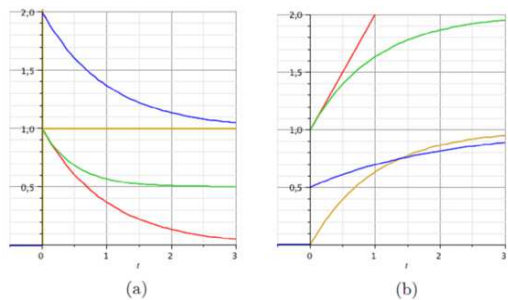}%
\caption{(a) Respostas do m\'{o}dulo de relaxa\c{c}\~{a}o da tens\~{a}o para
os modelos de Maxwell, Voigt, Maxwell modificado e Voigt modificado. (b)
Resposta da flu\^{e}ncia da deforma\c{c}\~{a}o para os respectivos modelos.}%
\end{figure}

O leitor pode n\~{a}o ter percebido, mas observamos um aspecto interessante a
respeito dos dois casos ideais limites: a lei de Hooke idealizada pela
Eq.(\ref{Hooke}) nos diz que a tens\~{a}o $\sigma(t)$ \'{e} diretamente
proporcional \`{a} \emph{derivada de ordem zero} da deforma\c{c}\~{a}o
$\epsilon(t)$; enquanto a lei de Newton idealizada pela Eq.(\ref{Newton}) nos
diz que a tens\~{a}o $\sigma(t)$ \'{e} diretamente proporcional \`{a}
\emph{derivada de ordem um} da deforma\c{c}\~{a}o $\epsilon(t)$. Ent\~{a}o uma
generaliza\c{c}\~{a}o para o comportamento viscoel\'{a}stico dos materiais,
segundo a ideia de derivadas de ordem fracion\'{a}ria, seria conjecturar que a
tens\~{a}o $\sigma(t)$ \'{e} diretamente proporcional a \emph{derivada
fracion\'{a}ria de ordem }$\alpha$, sendo $0<\alpha<1$, segundo a
equa\c{c}\~{a}o do tipo\footnote{Este modelo \'{e} conhecido como modelo de
Scott-Blair \cite{Mainardi}.}%
\begin{equation}
\sigma(t)=E\tau^{\alpha}\mathbf{D}^{\alpha}\epsilon(t),\text{ }\tau=\frac
{\eta}{E},\label{Scott-Blair}%
\end{equation}
sendo $\eta$ e $E$ os coeficientes de viscosidade e el\'{a}stico do material,
respectivamente. Claramente os casos ideais limites s\~{a}o recuperados quando
tomamos $\alpha=0$ (Hooke) ou $\alpha=1$ (Newton). O operador $\mathbf{D}%
^{\nu}$ seria qualquer operador de diferencia\c{c}\~{a}o fracion\'{a}ria
escolhido para ser usado (e.g., Riemann-Liouville, Caputo, etc...). Em
parti-cular, se quisermos um modelo que inclua todo o hist\'{o}rico do
material (em fun\c{c}\~{a}o do tempo), podemos estabelecer que o limite
inferior seja $t=-\infty$ e ainda, exigindo a causalidade do sistema (i.e.,
que as fun\c{c}\~{o}es sejam identicamente nulas para $t\leq0$), ent\~{a}o as
defini\c{c}\~{o}es segundo Riemann-Liouville e Caputo coincidiriam.

As fun\c{c}\~{o}es de respostas (m\'{o}dulo de relaxa\c{c}\~{a}o
$\mathbf{G}_{SB}(t)$ e flu\^{e}ncia $\mathbf{J}_{SB}(t)$) relacionadas ao
mo-delo da Eq.(\ref{Scott-Blair}) podem ser obtidas, por exemplo, via a
metodologia da transformada de Laplace como discutido na se\c{c}\~{a}o
anterior e resultam em%
\begin{align}
\mathbf{G}_{SB}(t)  &  =\frac{E\tau^{\alpha}}{\Gamma\left(  1-\nu\right)
}t^{-\alpha},\label{SBG}\\
\mathbf{J}_{SB}(t)  &  =\frac{1}{E\tau^{\alpha}\Gamma\left(  1+\nu\right)
}.\label{SBJ}%
\end{align}

Estudos comprovam que mesmo este modelo simplista, resulta em
descri\c{c}\~{o}es qualitativas mais precisas do comportamento de materiais
reais quando submetidos aos mesmos testes descritos na se\c{c}\~{a}o anterior
e mesmo as descri\c{c}\~{o}es quantitativas parecem mais promissoras. De fato,
podemos observar pela Figura 7, que todas as respostas (m\'{o}dulo de
relaxa\c{c}\~{a}o $\mathbf{G}_{SB}(t)$ e flu\^{e}ncia $\mathbf{J}_{SB}(t)$)
para diversas ordens $\alpha$ se comportam quantitativamente como o esperado,
i.e., o m\'{o}dulo de relaxa\c{c}\~{a}o possui um valor infinitamente grande
para $t=0$ e decaindo para relaxamento total quando $t\rightarrow\infty$,
j\'{a} a flu\^{e}ncia mostra o crescimento da deforma\c{c}\~{a}o de forma mais
apropriada com os experimentos mantendo a continuidade em $t=0$, lembrando que
os gr\'{a}ficos dados na Figura 7 foram feitos a partir de excita\c{c}\~{o}es
modeladas pela fun\c{c}\~{a}o degrau $H(t)$.%

\begin{figure}[ptb]%
\centering
\includegraphics[
height=1.5385in,
width=2.6602in
]%
{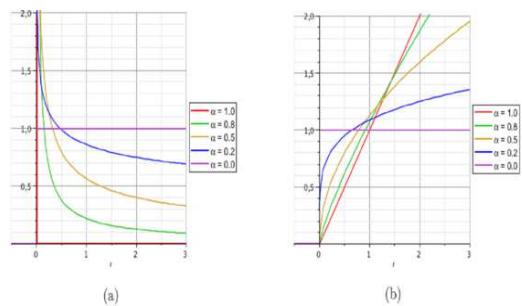}%
\caption{Respostas de m\'{o}dulo de \ relaxa\c{c}\~{a}o para o modelo
fracion\'{a}rio de Scott-Blair para diferentes par\^{a}metros $\alpha$,
($E=1$, $\eta=1$); (b) Respostas da flu\^{e}ncia da deforma\c{c}\~{a}o para o
modelo fracion\'{a}rio de Scott-Blair para diferentes par\^{a}metros $\alpha$
($E=1$, $\eta=1$).}%
\end{figure}

Este exemplo, mesmo que simples, mostra o potencial do uso do c\'{a}lculo
fracion\'{a}rio para abordarmos o problema de viscoelasticidade linear e de
fato, este ramo tem se mostrado um terreno f\'{e}rtil para estudiosos da
\'{a}rea com diversos livros \cite{Hilfer, Mainardi, Podlubny} mostrando
detalhamento dos modelos e resultados experimentais.

Mencionamos tamb\'{e}m outras \'{a}reas em que o c\'{a}lculo fracion\'{a}rio
tem se mostrado uma ferramenta poderosa, a saber, na descri\c{c}\~{a}o de
osciladores harm\^{o}nicos fracion\'{a}rios, for\c{c}as de fric\c{c}\~{a}o
fracion\'{a}rias, controladores fracion\'{a}rios, equa\c{c}\~{o}es de ondas
fracion\'{a}rias, etc... \cite{Herrmann, Mainardi, Podlubny, Shantanu,
Tarasov1}.

\subsection{Osciladores Harm\^{o}nicos\label{OscHarm}}

Nesta se\c{c}\~{a}o discutimos uma poss\'{\i}vel genera-liza\c{c}\~{a}o do
problema associado ao oscilador harm\^{o}nico. Recuperamos o caso inteiro e a
esse propomos uma generaliza\c{c}\~{a}o no sentido fracion\'{a}rio, isto
\'{e}, introduzimos dois par\^{a}metros associados \`{a} ordem da
equa\c{c}\~{a}o fracion\'{a}ria. A fim de obtermos uma solu\c{c}\~{a}o dessa
equa\c{c}\~{a}o, utilizamos a metodologia da transformada de Laplace e
fornecemos essa solu\c{c}\~{a}o expl\'{\i}cita para o caso em que n\~{a}o
temos o termo associado com o atrito, ou seja, consideramos o oscilador sem
amortecimento e apresentamos essa solu\c{c}\~{a}o em termos das
fun\c{c}\~{o}es de Mittag-Leffler.

Vamos considerar apenas o caso em que a derivada \'{e} tomada no sentido de
Caputo (\textbf{Defini\c{c}\~{a}o \ref{DFC}}) enquanto as condi\c{c}\~{o}es
iniciais admitem a cl\'{a}ssica interpreta\c{c}\~{a}o. Para discutirmos tal
problema inicia-se por obter uma equa\c{c}\~{a}o integral correspondente \`{a}
equa\c{c}\~{a}o diferencial associada ao oscilador harm\^{o}nico
\[
\frac{\mathbf{d}^{2}}{\mathbf{d}t^{2}}x(t)+\mu\frac{\mathbf{d}}{\mathbf{d}%
t}x(t)+\omega^{2}x(t)=0
\]
com $\mu\geq0$, representando o termo associado ao atrito e $\omega>0$ \'{e} a
frequ\^{e}ncia do oscilador, sendo as condi\c{c}\~{o}es iniciais $x(0)$ e
$x^{\prime}(0)$ dadas.

Integrando essa equa\c{c}\~{a}o diferencial duas vezes podemos escrever
\begin{align*}
x(t)  &  =x(0)+\mu t\,x(0)+t\,x^{\prime}(0)-\\
&  \mu\int_{0}^{t}x(v)\mathbf{d}v-\omega^{2}\int_{0}^{t}\int_{0}%
^{v}x(u)\mathbf{d}u\,\,\mathbf{d}v
\end{align*}
que, ap\'{o}s a utiliza\c{c}\~{a}o do teorema de Goursat \cite{Kilbas, Miller}
permite escrever
\begin{align*}
x(t)  &  =x(0)+\mu t\,x(0)+t\,x^{\prime}(0)-\\
&  \mu\int_{0}^{t}x(v)\mathbf{d}v-\omega^{2}\int_{0}^{t}(t-u)x(u)\,\mathbf{d}%
u.
\end{align*}
Ent\~{a}o, dados $x(0)$ e $x^{\prime}(0)$ obtemos uma equa\c{c}\~{a}o integral
associada ao problema do oscilador harm\^{o}nico equivalente ao problema
composto pela equa\c{c}\~{a}o diferencial e as condi\c{c}\~{o}es iniciais.

A fim de explicitar os c\'{a}lculos, consideremos o oscilador harm\^{o}nico
livre, isto \'{e}, sem atrito com a equa\c{c}\~{a}o integral j\'{a} na forma
fracion\'{a}ria
\begin{equation}
x(t)=x(0)+t\,x^{\prime}(0)-\frac{\omega^{2}}{\Gamma(\alpha)}\int_{0}%
^{t}(t-u)^{\alpha-1}x(u)\,\mathbf{d}u\label{EI}%
\end{equation}
com $1<\alpha\leq2$. Escolhemos esse intervalo a fim de recuperar o resultado
do oscilador harm\^{o}nico cl\'{a}ssico, no caso em que $\alpha=2$.

Note que essa equa\c{c}\~{a}o integral fracion\'{a}ria corresponde \`{a}
equa\c{c}\~{a}o diferencial fracion\'{a}ria
\[
_{0}^{C}\mathcal{D}_{x}^{\alpha}x(t)+\omega^{\alpha}x(t)=0
\]
e as condi\c{c}\~{o}es iniciais $x(0)$ e $x^{\prime}(0)$ sendo $1<\alpha\leq2
$ e a derivada considerada no sentido de Caputo. (\textbf{Defini\c{c}\~{a}o
\ref{DFC}}).

Vamos procurar uma solu\c{c}\~{a}o da equa\c{c}\~{a}o integral atrav\'{e}s da
metodologia da transformada de Laplace que ap\'{o}s aplicada na Eq.(\ref{EI})
e utilizando a defini\c{c}\~{a}o do produto de convolu\c{c}\~{a}o permite
escrever%
\[
F(s)=\frac{x(0)}{s}+\frac{x^{\prime}(0)}{s^{2}}-\omega^{\alpha}\int%
_{0}^{\infty}\frac{t^{\alpha-1}}{\Gamma(\alpha)}\star x(t)\,{\mbox{e}}%
^{-st}\mathbf{d}t
\]
onde o s\'{\i}mbolo $\star$ denota o produto de convolu\c{c}\~{a}o. Logo,
obtemos
\[
F(s)=\frac{x(0)}{s}+\frac{x^{\prime}(0)}{s^{2}}-\frac{\omega^{2}}{s^{\alpha}%
}F(s)
\]
sendo
\[
F(s)=\int_{0}^{\infty}x(t)\,{\mbox{e}}^{-st}\,\mathbf{d}t
\]
a transformada de Laplace de $x(t)$ de par\^{a}metro $s$ com
$\operatorname{Re}(s)>0$. Isolando $F(s)$ podemos escre-ver, j\'{a}
rearranjando,
\[
F(s)=x(0)\frac{s^{-1}}{1+\omega^{\alpha}s^{-\alpha}}+x^{\prime}(0)\frac
{s^{-2}}{1+\omega^{\alpha}s^{-\alpha}}.
\]

Para recuperar a solu\c{c}\~{a}o $x(t)$ tomamos a transformada de Laplace
inversa. Logo para
\[
x(t)=\frac{1}{2\pi i}\int_{c-i\infty}^{c+i\infty}{\mbox{e}}^{st}%
F(s)\,\mathbf{d}s\equiv%
\mathcal{L}%
^{-1}[F(s)],
\]
obtemos a solu\c{c}\~{a}o na forma%
\[
x(t)=x(0)\,E_{\alpha}(-\omega^{\alpha}t^{\alpha})+x^{\prime}(0)\,E_{\alpha
,2}(-\omega^{\alpha}t^{\alpha})
\]
onde $E_{\alpha}(\cdot)$ e $E_{\alpha,\beta}(\cdot)$ s\~{a}o as
fun\c{c}\~{o}es de Mittag-Leffler com um e dois par\^{a}metros,
respectivamente \cite{Kilbas, Mainardi, Gorenflo}. Admita, enfim, que $x(0)=1$
e $x^{\prime}(0)=0 $ logo a solu\c{c}\~{a}o neste caso \'{e}
\[
x(t)=E_{\alpha}(-\omega^{\alpha}t^{\alpha})
\]
que, no caso extremo $\alpha=2$ fornece $x(t)=\cos\omega t$ que \'{e} a
solu\c{c}\~{a}o do problema associado ao oscilador harm\^{o}nico de ordem inteira.

\section{Conclus\~{a}o}

Neste trabalho foram introduzidas as chamadas equa\c{c}\~{o}es diferenciais
fracion\'{a}rias e constatado que a metodologia da transformada de Laplace
mostra-se, assim como no caso das EDOs de ordem inteiras com coeficientes
constantes, uma \'{o}tima ferramenta a ser usada em vista dos benef\'{\i}cios
da sua simplicidade. Em particular, verificou-se de forma indireta que da
mesma forma que a fun\c{c}\~{a}o exponencial $e^{x}$ desempenha um papel de
destaque no c\'{a}lculo cl\'{a}ssico de ordem inteira, as fun\c{c}\~{o}es de
Mittag-Leffler desempenham um papel destacado an\'{a}logo no c\'{a}lculo de
ordem arbitr\'{a}ria. O que queremos denotar por esta afirma\c{c}\~{a}o \'{e}
que assim como se faz necess\'{a}rio um bom entendimento da fun\c{c}\~{a}o
exponencial para o estudo e aplica\c{c}\~{o}es do c\'{a}lculo diferencial e
integral cl\'{a}ssicos o mesmo pode ser dito \`{a} respeito das
fun\c{c}\~{o}es de Mittag-Leffler para aqueles que desejam se aventurar pelo
c\'{a}lculo fracion\'{a}rio. De fato, n\~{a}o \'{e} incomum ao resolvermos
equa\c{c}\~{o}es diferenciais (ordin\'{a}rias) de ordem fracion\'{a}ria,
obtermos solu\c{c}\~{o}es que, se n\~{a}o diretamente expressas em termos de
uma fun\c{c}\~{a}o de Mittag-Leffler, ent\~{a}o ao menos se relacionam com ela
atrav\'{e}s de alguma identidade.

Por fim, observamos que, de modo relativamente simples, problemas
cl\'{a}ssicos da F\'{\i}sica-Matem\'{a}tica ficam bem modelados sob a
\'{o}ptica do c\'{a}lculo fracion\'{a}rio e d\~{a}o resultados promissores
(inclusive coincidindo com os resultados de ordem inteira j\'{a} firmados)
mesmo quando s\~{a}o elaborados de forma simples, como verificados nos
exemplos de viscoelasticidade linear e osciladores harm\^{o}nicos.

O leitor interessado em conhecer mais modelagens usando o c\'{a}lculo
fracion\'{a}rio, pode recorrer as fontes \cite{Oldham, Abbas, Capelas1,
Capelas2, Cresson, Garra, Costa, Tenreiro, West}.

\bigskip

\textbf{AGRADECIMENTOS}

FGR agradece a CAPES pela bolsa dispon\'{\i}vel durante o programa de doutorado.

Os autores tamb\'{e}m s\~{a}o gratos ao \'{a}rbitro pelos pareceres e
sugest\~{o}es dadas para tornar este trabalho melhor enquadrado ao p\'{u}blico alvo.

\end{document}